%% file: Main_File.tex
\documentclass[journal]{IEEEtran}
\usepackage{cite}
\usepackage{algorithmic}
\usepackage{graphicx}
\usepackage{textcomp}
\usepackage{xcolor}
\usepackage{caption} 
\usepackage{bbm} 
\usepackage{steinmetz}
\usepackage{bm}
\usepackage{caption}
\usepackage{subcaption}
\usepackage{mathtools}

\usepackage{amsmath,amssymb,amsfonts}

\DeclareMathOperator*{\argmin}{arg\,min}
\usepackage{url}
\usepackage{fancyhdr}
\begin{document}

%
\title{Road State Inference via Channel State Information
}
%
%
%

\author{Halit Bugra~Tulay,~\IEEEmembership{Member,~IEEE,}
        Can Emre~Koksal,~\IEEEmembership{Senior Member,~IEEE}
\thanks{

Halit Bugra Tulay and Can Emre Koksal are with the Department of Electrical and Computer Engineering, The Ohio State University, OH, 43202 USA (e-mail: tulay.1@osu.edu; koksal.2@osu.edu) }
\thanks{\hfill}

}

\maketitle
\thispagestyle{fancy}

\begin{abstract}
A wide variety of sensor technologies are recently being adopted for traffic monitoring applications. Since most of these technologies rely on wired infrastructure, the installation and maintenance costs limit the deployment of the traffic monitoring systems. In this paper, we introduce a traffic monitoring approach that exploits physical layer samples in vehicular communications processed by machine learning techniques. We verify the feasibility of our approach with extensive simulations and real-world experiments. First, we simulate wireless channels under realistic traffic conditions using a ray-tracing simulator and a traffic simulator. Next, we conduct experiments in a real-world environment and collect messages transmitted from a roadside unit (RSU). The results show that we are able to predict different levels of service with an accuracy above 80\% both on the simulation and experimental data. Further, the proposed approach is capable of estimating the number of vehicles with a low mean absolute error on both data. Our approach is suitable to be deployed alongside the current monitoring systems. It doesn't require additional investment in infrastructure since it relies on existing vehicular networks.
\end{abstract}

\begin{IEEEkeywords}
traffic monitoring, intelligent transportation systems, DSRC, C-V2X, vehicular ad-hoc networks
\end{IEEEkeywords}

%
\IEEEpeerreviewmaketitle

\input{introduction.tex}
\input{related_work.tex}
\input{problems.tex}

\input{approach.tex}

\input{performance.tex}
\input{conclusion.tex}

\bibliography{bibtex/bib/IEEEexample.bib}{}
\bibliographystyle{IEEEtran}

\begin{IEEEbiography}[{\includegraphics[width=1in,height=1.25in,clip,keepaspectratio]{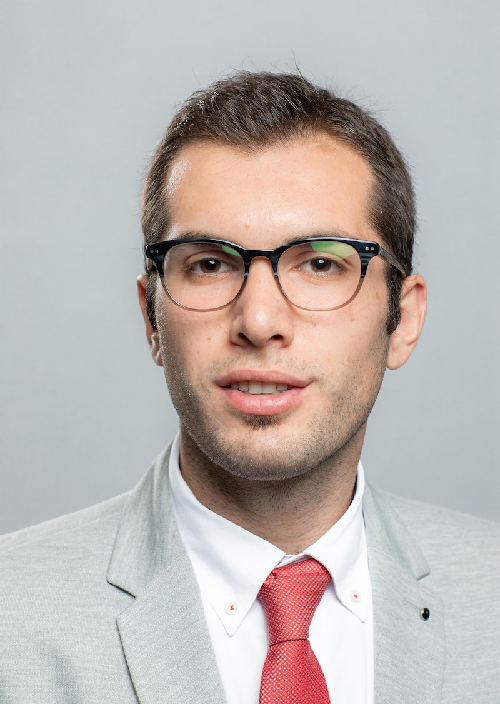}}]{Halit Bugra Tulay} (S’19) received the B.S. degree in the electrical and electronics engineering from
Hacettepe University in 2016. He is currently pursuing the Ph.D. degree in the
department of electrical and computer engineering at The Ohio State University.
His research interests include wireless communication, cybersecurity and machine learning.
\end{IEEEbiography}

\begin{IEEEbiography}[{\includegraphics[width=1in,height=1.25in,clip,keepaspectratio]{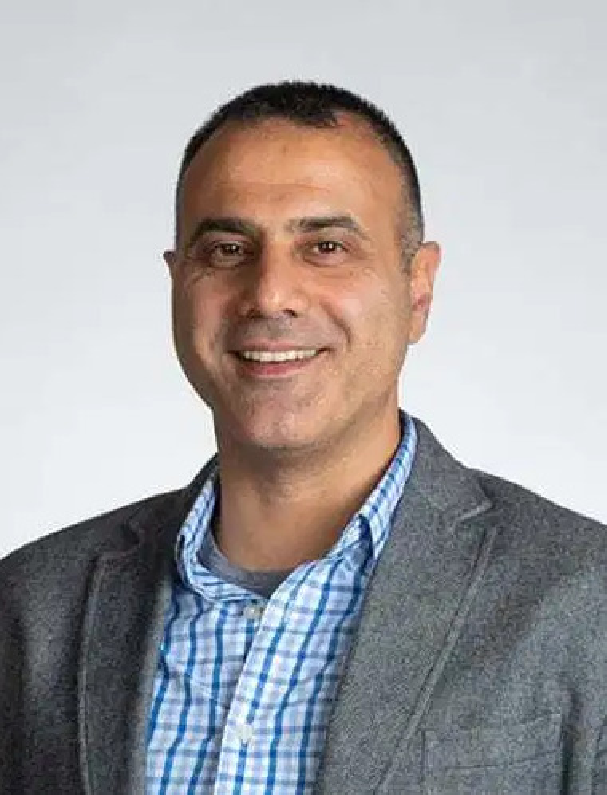}}]{Can Emre Koksal}
(S’96–M’03–SM’13) received the M.S. and Ph.D. degrees
in electrical engineering and computer science from MIT, in 1998 and 2002,
respectively. He is professor at the Electrical and
Computer Engineering Department of The Ohio State University since 2006. He researches on wireless communication, cybersecurity, information theory, stochastic processes.
He served as an Associate Editor for the IEEE Transactions on Information Theory, the IEEE Transactions on Wireless Communications, and Computer Networks.
\end{IEEEbiography}





\end{document}

%% file: introduction.tex
\section{INTRODUCTION}
A traffic monitoring system (TMS) is used to collect traffic data such as traffic density, types of vehicles, and speed to perform traffic analysis, predict future transportation needs, and improve the safety of transportation based on the collected data. Nowadays, intrusive sensors (e.g., inductive loops, magnetic detectors), vision-based systems, or radars are mostly used for traffic monitoring. Most of these systems need fixed wired infrastructure which results in high installation and maintenance costs. Consequently, the costs of the systems prevent the dense deployment on roads. According to the Georgia Department of Transportation, a TMS on two-lane roadway costs roughly \$25,000 and the cost of installation can go up to \$80,000 \cite{GDOT}. This motivates low-cost, non-intrusive approaches.  

\begin{figure}[t]
\includegraphics[width=0.5\textwidth]{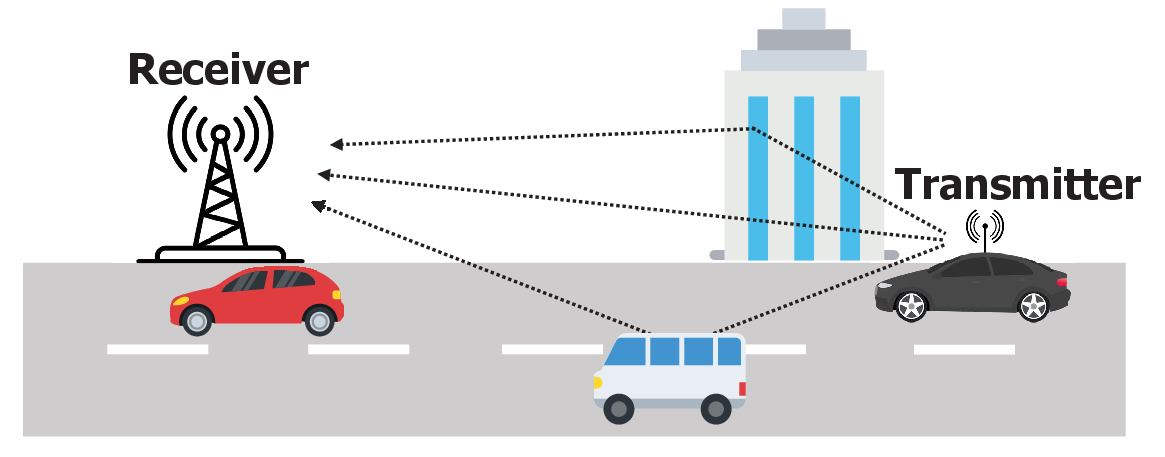}
\centering
 \caption{\label{traffic} A typical communication scenario in a vehicular network. }
\end{figure}

In this paper, we introduce a novel traffic monitoring approach that exploits the communication signals broadcast in a vehicular ad-hoc network (VANET). We solely rely on the channel state information, not the content of the messages. In VANETs, vehicles and traffic infrastructure exchange data periodically with each other via vehicle-to-infrastructure (V2I) communication links. The vehicles equipped with onboard units (OBUs) exchange data such as their position, speeds, and certain event-triggered messages with the other vehicles and roadside units (RSUs) in the traffic infrastructure. Similarly, RSUs inform vehicles about signal phases of traffic lights, speed limits, or road work maps\cite{soto2022survey}. We utilize the signals transmitted from a vehicle and captured at deployed roadside units (RSUs) as shown in Fig. \ref{traffic}, or vice-versa. Our approach doesn't require additional hardware deployment while relying on the existing infrastructure. We also don't rule out the possibility of static transmitters or receivers placed on the side of a roadway. Since our approach is based on the use of mere channel state information over a single RF chain, it can be an effective candidate across different standards, including DSRC \cite{DSRC_Stand} and cellular vehicle-to-everything (C-V2X) which is defined by the 3GPP as part of its LTE and ongoing 5G families of standards. We also demonstrate that our method provides highly accurate results in sparse settings, even when there is a single transmitter-receiver pair. This may be highly valuable until we transition to dense deployments of V2X systems.

Our approach takes advantage of the radio propagation characteristics of vehicular networks. A transmitted wireless signal travels over multiple paths reflected from the surfaces. In vehicular networks, reflectors are mostly vehicles on a road and the channel state is shaped by the traffic conditions. The idea in our approach is to infer the traffic conditions on the road based on the channel state information, specifically the channel frequency response (CFR). We use supervised learning algorithms since a structured model-based mapping between the traffic conditions and the channel frequency response values is extremely difficult due to many possible traffic conditions. Unfortunately, there is no available signal-level data collected under a variety of traffic conditions to train the machine learning algorithms. To solve this problem, we use a ray-tracing simulator and a traffic simulator in conjunction to simulate the wireless propagation for a four-lane road. We first train the machine learning algorithms with the obtained simulation data and evaluate the performance of the proposed approach. Subsequently, we conduct DSRC-based V2I experiments in a real vehicular network at an intersection. Specifically, we collect signal phase and timing (SPaT) messages broadcast by a real roadside unit (RSU). The messages have been collected from a static receiver and a moving vehicle using our software-defined radio, and we evaluate the performance with the collected experimental data.

Operational conditions in a traffic stream are characterized by quantitative measures specified in the highway capacity manual \cite{manual2000highway}. The highway capacity manual defines six levels of service (LOS), from letter A to F, for the density of traffic in terms of passenger cars per kilometer per lane while LOS A represents the best operating conditions and LOS F represents the worst conditions. Our approach estimates the LOS with over 80\% classification accuracy on both the simulation data and the experimental data. We have estimated the number of vehicles with a mean absolute error (MAE) of 2.74 vehicles on a four-lane road in the simulations, and with a mean absolute error of 0.93 vehicles on a two-lane road in the real-world experiments. The main contribution of this paper can be summarized as:

\begin{itemize}
  \item We propose a non-intrusive and cost-effective solution for traffic inference by exploiting the existing VANET infrastructure.
  \item We demonstrate the viability of the use of mere channel frequency responses within the existing vehicular communication technologies for high-performance traffic state inference, without requiring any additional hardware deployment.
  \item We collected a substantial amount of V2I communication data from a roadside unit and also generated simulation data using a ray-tracing simulator. The datasets are available at \cite{dataset} for the research community use.
  \item We obtain multiple trained models with different machine learning algorithms, and we show that they are able to map the CFR values to traffic conditions with a low error.
\end{itemize}

This paper extends our previous work \cite{VTC2020} by increasing the number of classes for the level of service prediction, using data preprocessing techniques, and with extended real-world measurements. This paper is organized as follows. In Section II, we briefly discuss the previous work done related to the topic and Section III describes the problem model. Section IV describes the details of our approach which include data creation, data preprocessing, and machine learning algorithms. In Section V, we demonstrate the experimental and simulation results. Finally, Section VI summarizes our work.

%% file: related_work.tex
\section{Related Work}\label{related_work}

There has been extensive research on traffic monitoring/management systems, and different approaches have been proposed \cite{nellore2016survey} \cite{de2017traffic}. The operation of a traffic monitoring system depends on various technologies. According to the used technology, we can categorize these systems into three groups: 1) Sensor-based systems 2) Aerial technology-based systems and 3) Wireless technology-based systems.

\textbf{Sensor-based Systems:} There is a wide variety of sensors used in traffic monitoring systems today. These sensors can be classified as intrusive and non-intrusive sensors. Intrusive sensors are installed directly into the road surface. Sensors such as inductive loops, magnetic detectors, and other weigh-in-motion devices constitute the popular intrusive sensors. For example, the freeway performance measurement system (PeMS) is a system used by the California department of transportation (Caltrans). It is based on 30-second measurements from inductive loops in real time. The data comprises the number of vehicles crossing the loop and occupancy (the average fraction of time a vehicle is present over the loop). PeMS data can be accessed online \cite{pems}. PeMS achieves an average accuracy of 82\% on the level of service classification task, while it is as low as 56\% for LOS E \cite{khan2017real}. The major drawback of the intrusive sensors is the interruption of traffic since they are installed under road surfaces. Also, the cost of the system increases if the sensor is able to monitor a single lane while the multiple lanes are to be monitored \cite{mimbela2007summary}. 

Non-intrusive sensors are deployed above the road level or on the side of a roadway. As a result, they can be easily deployed, and the installation of the systems does not interrupt traffic flow. Cameras, microwave radars, and passive infrared sensors are among the most popular non-intrusive sensors. Currently, many traffic monitoring systems incorporate cameras and video processing techniques \cite{vision-based_monitoring}. As deep learning techniques are becoming more widely adopted, they offer great potential for traffic monitoring applications over traditional techniques. In \cite{pamula2018road}, the authors evaluate the performance of different convolutional neural network (CNN) architectures on the level of service classification. They achieve a mean accuracy of more than 80\% for tested CNN configurations, while the best architecture reaches a mean accuracy of 89\%. Despite the wide usage of non-intrusive sensors, their deployment and maintenance costs are relatively high. Further, their performance can be affected by certain weather conditions \cite{middleton2007state}.

\textbf{Aerial technology-based systems:} The aerial platforms, especially the unmanned aerial vehicles (UAVs), have become a cost-efficient solution for road traffic monitoring because of their mobility and large range \cite{kanistras2013survey}. These technologies don't require any hardware deployment under/on the roadways, and therefore, they are suitable for such a dynamic environment. In \cite{zhao2017automated}, the authors present a computer vision-based traffic surveillance system using an aerial camera array. They present that the deep learning combined with speeded up robust features (SURF)-based approach is able to achieve over 93\% accuracy in density estimation.

\textbf{Wireless technology-based systems:} The U.S. Department of Transportation's intelligent transportation systems (ITS) Joint Program Office encourages to deploy applications utilizing data captured from multiple sources (e.g., vehicles, mobile devices, and infrastructure) across all elements of the surface transportation systems \cite{its_program}. To this end, connected vehicle technologies, DSRC and C-V2X, have been considered an important part of intelligent transportation systems. Architecture Reference for Cooperative and Intelligent Transportation (ARC-IT) \cite{arc-it} of the US Department of Transportation uses the collected data from connected vehicles to estimate traffic conditions. To this end, traffic monitoring using vehicle-to-vehicle (V2V) and vehicle-to-infrastructure communication links has been studied in several studies. In \cite{khan2017real}, the authors propose a method for estimating traffic density using connected vehicle technology and artificial intelligence. Using VISSIM, a traffic microsimulation software, they demonstrate that the accuracy of the LOS classification is a minimum of 85\% with 20\% and greater connected vehicle penetration levels. In \cite{bauza2013traffic}, the authors introduce CoTEC (COperative Traffic congestion detECtion), a road congestion detection algorithm based on cooperative awareness messages (CAM) or beacon messages broadcasting the road traffic conditions periodically. CoTEC is evaluated under large-scale highway scenarios using iTETRIS, an open-source simulation platform created to investigate the impact of cooperative vehicular systems. In \cite{scorpion}, the authors propose a method named SCORPION (System with COoperative Routing to imProve traffIc cONdition) that is based on V2I communication. They use the K-Nearest Neighbor (KNN) classifier which uses the average speed and the density of each road to classify the traffic conditions as free-flow, slightly congested, moderately congested, and severely congested.

Recent advances in wireless technology also enable a new sensing paradigm that is often used to recognize human behaviors using WiFi technologies in indoor settings \cite{ma2019wifi}. Similarly, WiFi-based traffic monitoring systems have been proposed with the motivation of reducing deployment costs for large-scale deployment. The researchers study the feasibility of using WiFi devices for traffic monitoring applications using channel state information (CSI) \cite{won2017witraffic}\cite{won2019deepwitraffic}, received signal strength indicator (RSSI) \cite{haferkamp2017radio} \cite{gupta2018low}, link quality indicator (LQI) \cite{roy2011wireless}, and packet loss rate \cite{roy2011wireless}. The proposed approaches exploit the RF propagation between a receiver and a transmitter that is affected by passing vehicles. Consequently, different patterns of the wireless channel metrics are observed at the receiver depending on traffic conditions and the types of vehicles. These patterns are used to classify and count passing vehicles. The system proposed in \cite{won2017witraffic} utilizes a laptop and a router deployed on the roadside. The CSI powers of the passing vehicles are captured on local roads and highways to count and classify vehicles using machine learning techniques. They show that road lanes have different CSI patterns and this allows them to identify in which lane a vehicle is detected. A similar setup and deep learning techniques are used in their next study\cite{won2019deepwitraffic} to count and classify vehicles. In \cite{haferkamp2017radio}, the authors employ a system that exploits the attenuation of radio signals for the detection and classification of vehicles using machine learning techniques. Similarly, the authors in \cite{gupta2018low} proposed a system that exploits RSSI information to detect and classify vehicles. The authors in \cite{roy2011wireless} set up a transmitter-receiver pair and classify traffic conditions as free-flow or congested using a decision tree-based classifier. In the proposed method, receivers collect packets from the transmitter placed on the opposite side of a road and different metrics (RSSI, LQI, packet loss) are used to classify the traffic conditions. They achieve a classification accuracy of 97\% in their experiments.

The aforementioned approaches require specialized hardware (e.g., inductive loops, UAVs, camera arrays). In contrast, our approach relies on the samples collected in available infrastructure without a need to decrypt the message content, thereby preserving privacy as a bonus. Also, our approach is non-intrusive, easy to deploy, and cost-effective since it is built upon the existing infrastructures. Thus, it can easily be integrated with other traffic monitoring systems and improve their performance. A unique component of our approach is that it doesn't observe the wireless link behavior (interruption of line-of-sight component) for counting each vehicle as opposed to the previous approaches. We estimate the number of vehicles from a single channel frequency estimate. Another unique component of our work is in the measurements. We rely heavily on actual data collected from a real roadside unit, as opposed to analysis or raw simulations.

%% file: problems.tex
\section{PROBLEM MODEL AND STATEMENT}\label{problem_section}

\begin{figure}[h]
\includegraphics[width=0.5\textwidth]{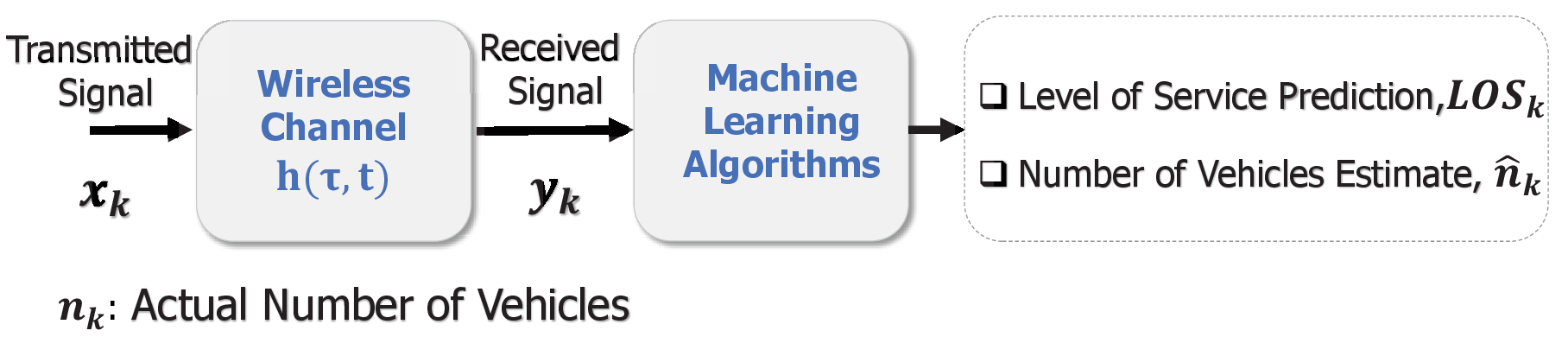}
\centering
 \caption{\label{problem} Problem model.}
\end{figure}

The components of the system are shown in Fig. \ref{problem}. $x_k$ are transmitted signal samples of $k^{th}$ frame from a static transmitter on the roadside or a vehicle, and $y_k$ are the received signal samples after passing through the wireless channel. The receiver estimates the wireless channel using the received signal samples. We can represent the wireless channel as a linear time-varying channel filter and it can be characterized by its baseband impulse response as:
\begin{equation}
    h_b(t,\tau)=\sum_i a_i(t)e^{-j2\pi f_c \tau_i(t)}g(\tau-\tau_i(t))
\end{equation}
where $a_i(t)$, $\tau_i(t)$, $f_c$  are the path attenuation and delay of path $i$ at time $t$, the carrier frequency, respectively. $g(\tau)$ is the impulse response of the transmit and receiver filters. The corresponding time-varying channel frequency response can be calculated as \cite{tse2005fundamentals}:
\begin{equation}
\label{CFR_equation}
    \small
    H(t,f) = \int\limits_{-\infty}^\infty h_b(\tau,t) e^{-j2\pi f \tau} \, d\tau= G(f)\sum_i a_i(t) e^{-j2\pi (f+f_c) \tau_i(t)}
\end{equation}
where $G(f)$ is the frequency response of the transmit and receive filters and it can be assumed to be constant in the presence of guard subcarriers on both sides of the spectrum \cite{6814271}. We consider an orthogonal frequency division multiplexing (OFDM) system. The channel frequency response is estimated using the preambles at the beginning of each OFDM frame. We can estimate the channel frequency response (CFR) at $N$ subcarriers for $k^{th}$ frame as:
\begin{equation}
    \mathbf{H_k} \triangleq [H_{k,0},\  H_{k,1},\ \dots,\ H_{k,N-1}]^T
\end{equation}  
where $H_{k,i} \triangleq H(kT,f_i)$. T denotes the frame duration and $f_i$ is the baseband frequency of $i^{th}$ subcarrier. Note that $H_{k,i}$ is a complex number and represented by the magnitude $|H_{k,i}|$ and the phase $\angle H_{k,i}$ as $H_{k,i}=|H_{k,i}|e^{j\angle H_{k,i}}$. Considering the phase estimation errors and the phase noise, we only use the magnitude frequency response values, $|H_{k,i}|$, estimated from the received signal. The machine learning algorithms are trained with the magnitude response values and output the level of service prediction, $LOS_k$, and estimated number of vehicles, $\hat{n}_k$.

\textbf{Problem Statement:} 
First, we predict the density of the traffic conditions and assign level of service labels $LOS_k \in \{\text{A, B, C, D, E, F}\}$ for each received frame. The highway capacity manual defines six levels of service according to the type of roadway. In this work, we use the level of service definitions that are based on density. The density of a roadway is expressed in terms of passenger cars per kilometer per lane (pc/km/ln). Table \ref{los} summarizes the level of service thresholds based on density for a basic freeway segment. The machine learning algorithms map N-dimensional CFR vectors to level of service predictions, ${c}(|\mathbf{H}_k|): \mathbb{R}^{N} \xrightarrow{} LOS_k $, where $|\mathbf{H}_k|$ is the magnitude of CFR values for $k^{th}$ received frame and $c()$ represents the classification algorithm. 
\begin{table}[h]
\caption{\label{los} Level of service definitions according to density}
\begin{tabular}{|c|c|}
\hline
\textbf{Level of service} & \textbf{Density, pc/km/ln} \\ \hline
           A      &          0-7         \\ \hline
           B      &          7-11         \\ \hline
           C      &         11-16          \\ \hline
           D      &         16-22          \\ \hline
           E      &         22-28          \\ \hline
           F      &         $>$ 28          \\ \hline
\end{tabular}
\centering
\end{table}

Next, we estimate the number of vehicles on a roadway. In this problem, a regression algorithm maps the CFR vector to the number of vehicles estimate, $r(|\mathbf{H}_k|): \mathbb{R}^{N} \xrightarrow{} \mathbb{R}$, where r() represents the regression algorithm. To this end, we build a regression model that minimizes the error between the actual number of vehicles, $n_k$, and the estimated number of vehicles, $\hat{n}_k$.

%% file: approach.tex
\section{Approach}\label{approach_section}

We need training data that include the wireless channel realizations under different traffic conditions to train the machine learning algorithms. Unfortunately, such data are not available. Therefore, we first generate data by integrating two simulators. The integration of the simulators helps us to obtain realistic wireless data under complex scenarios easily and evaluate the performance of the approach. Subsequently, we collect real experimental data in different scenarios as described in the next sections.  

\subsection{Data Creation Methodology}

\subsection*{\textbf{Simulation of the Wireless Channel}}
In a vehicular network, vehicle surfaces, buildings, and terrains are the main sources of reflections and diffractions. The wireless channel between the highly mobile nodes is hard to be captured by the probabilistic channel models. In this work, we utilize the ray-tracing approach rather than probabilistic channel models. The ray-tracing techniques represent the electromagnetic waves sent from a transmitter as a simple particle and estimate the paths between a receiver and the transmitter. Therefore, ray-tracing provides more accurate and spatially consistent results compared to the probabilistic models. More details about the principles of ray tracing can be found in \cite{7152831}. In this work, we use Remcom's Wireless Insite \cite{wirelessinsite} as a ray-tracing simulator to simulate wireless propagation.

The ray-tracing simulator provides the rays, each corresponding to a propagation path, between a receiver and a transmitter. The simulator also gives information on power, delay, phase, angle of arrival, and angle of departure of each path. Given this information, we can calculate CFR values according to Eq. \ref{CFR_equation}. However, IEEE 802.11p utilizes a preamble-based channel estimation method from noisy received signals, and this results in an error in the channel estimation. We incorporate this error with the following model. Suppose that $\{t[n]\} _{n=0}^{N_{tr}-1}$ is a known training sequence, $\mathbf{y}=[$ y[L]  y[L+2] \dots y[$N_{tr}$-1] $]$ is the received signal samples, and $\mathbf{w}=[$ w[L]  w[L+2] \dots w[$N_{tr}$-1]$]$ is the noise samples after removing first L received samples. We can write the received signal in a matrix form as:
\begin{equation}
    \mathbf{y}_k=\mathbf{t}_k\mathbf{h}_k + \mathbf{w}_k
\end{equation}
where $\mathbf{t}_k$ is the $(N \times L)$ circularly shifted training matrix with $N=N_{tr} - L$ the received sequence length. $\mathbf{h}_k$ is an L-tap channel impulse response vector. The received frequency domain signal of the $k^{th}$ OFDM frame, after removing the cyclic prefix and applying the discrete Fourier transform, can be written in the vector form as $\mathbf{Y}_k\triangleq[Y_{k,0}, Y_{k,1}, \dots, Y_{k,N-1}]^T$ can be written as:
\begin{equation}
\label{freq_signal}
    \mathbf{Y}_k=\mathbf{T}_k\mathbf{H}_k+\mathbf{W}_k
\end{equation}
where $\mathbf{T}_k$ is $N\times N$ diagonal matrix with $<n,n>^{th}$ element given as $T_{k,n}$ where $T_{k,n}$ is the pilot subcarriers of $k^{th}$ frame. $\mathbf{H}_k$ is the channel frequency response, and $\mathbf{W}_k\triangleq[W_{k,0}, W_{k,1}, \dots, W_{k,N-1}]^T$ is the frequency domain noise vector where $W_{k,i}$ is an additive Gaussian noise at subcarrier $i$ of $k^{th}$ frame, with zero mean and variance $\sigma_W^2$. The commonly used channel estimation scheme for the IEEE 802.11p is the least-squares estimation and the least-squares channel estimate, which is also the maximum likelihood estimate under additive white gaussian noise (AWGN), is given by:
\begin{equation}
\label{ls_estimation}
    \begin{split}
    \hat{\mathbf{H}}_{k}&= \argmin_{\mathbf{H_k}} || \mathbf{Y_k} - \mathbf{T_k}\mathbf{H_k}|| ^2\\
    & = (\mathbf{T_k}^H \mathbf{T_k})^{\text{-1}} \mathbf{T_k}^H \mathbf{Y_k} = \mathbf{T_k}^{\text{-1}}\mathbf{Y_k}  \\
    & = \bigg[\frac{Y_{k,0}}{T_{k,0}}, \frac{Y_{k,1}}{T_{k,1}}, \dots, \frac{Y_{k,N-1}}{T_{k,N-1}} \bigg]  \\
    \end{split}
\end{equation}
where $||\cdot||$ denotes the norm of a vector, $x^H$ is Hermitian transposition of x. Substituting Eq. \ref{freq_signal} into Eq. \ref{ls_estimation}, the LS channel estimate of subcarriers can be expressed as:
\begin{equation}
    \begin{split}
    \hat{\mathbf{H}}_{k}= \mathbf{H}_k + \underbrace{\mathbf{T}_k^{\text{-1}}\mathbf{W}_k}_\text{$\mathbf{\mathcal{E}}$=Estimation Error}
    \end{split}
\end{equation}
where $\mathbf{\mathcal{E}}=\mathbf{T}_k^{\text{-1}}\mathbf{W}_k$ denotes the estimation error. Since $E[\hat{\mathbf{H}}_{k}] = E[ \mathbf{H}_k]+\mathbf{T}_k^{\text{-1}}E[\mathbf{W_p}]=E[ \mathbf{H}_k]$ forms an unbiased estimator of $\mathbf{H}_k$, i.e., $E[\mathbf{\mathcal{E}}]=\mathbf{0}$. The covariance matrix of the estimation error can be calculated as:
\begin{equation}
\label{simulation_errors}
    \begin{split}
    E[\mathcal{E} \mathcal{E}^H]
    & = E[\mathbf{T}_k^{-1}\mathbf{W_p} \mathbf{W_p}^H (\mathbf{T}_k^{-1})^H]\\
    & = \mathbf{T}_k^{-1} E[\mathbf{W}_k \mathbf{W}_k^H] (\mathbf{T}_k^{-1})^H\\
     & = \sigma_W^2 (\mathbf{T}_k^H \mathbf{T}_k)^{-1}\\
     & = \frac{\sigma_W^2}{P_T}  \mathbf{I}_N = \sigma_\mathcal{E}^2  \mathbf{I}_N
    \end{split}
 \end{equation}where $P_T$ is equal to the transmitter power per subcarrier (i.e., $(\mathbf{T_k}^H \mathbf{T_k})= P_T \mathbf{I_N}$, where $\mathbf{I}_N$ is an identity matrix of size N) and $\sigma_\mathcal{E}^2$ is the error variance per subcarrier. Note that the error variance per subcarrier depends on the noise variance, $\sigma_W^2$, and the transmitter power per subcarrier. 

We estimate the CFR values using the least-squares channel estimator in our real-world experiments. For simulations, the propagation data obtained from the ray-tracing simulator are used to generate the actual CFR values. Since the estimation error is not incorporated in our simulations, we calculate the estimation errors with Eq. \ref{simulation_errors} and add them to the actual CFR values.
\subsection*{\textbf{Simulation of the Traffic}}

Many models and simulation tools have been developed to generate realistic vehicular mobility \cite{silva2018survey}. In this work, we use Simulation of Urban Mobility (SUMO)\cite{SUMO2018} to simulate realistic traffic conditions for the ray-tracing simulator since it provides us great flexibility with various configuration files and allows us to import real roads from OpenStreetMap(OSM). The traffic control interface of SUMO, \textit{TraCI} \cite{TraCI} is also utilized to obtain different parameters of the vehicles such as vehicle ID, position, speed, etc., and change these parameters.

\subsection*{\textbf{Integration of the simulators}}

\begin{figure}[h]
\includegraphics[width=0.5\textwidth]{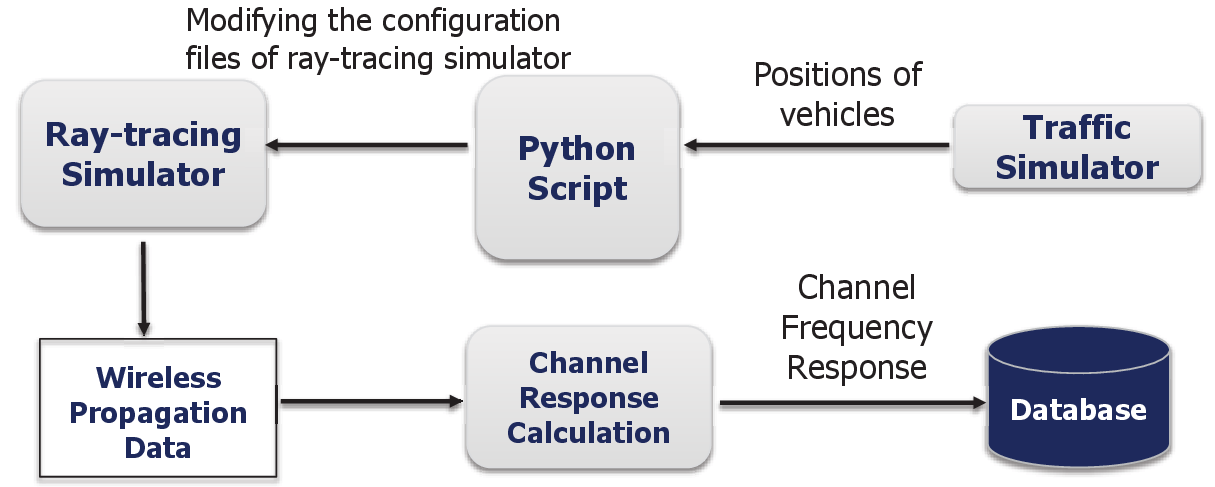}
\centering
\caption{\label{method}The methodology that integrates the simulators.}
\end{figure}

We follow the methodology proposed in \cite{tulay2021robust} as shown in Fig. \ref{method} to integrate two simulators and simulate the wireless propagation under various traffic conditions. In this methodology, the positions of vehicles obtained from the traffic simulator are used to place vehicles in the form of \textit{Objects} in the ray-tracing simulator. To enable this, we wrote a Python script that runs the traffic simulator for given mobility parameters and retrieves the position of vehicles. The script later modifies the positions of the objects in the configuration files of the ray-tracing simulator. 
After modifying the configuration files, the script uses the command line controls to run Wireless Insite's calculation engine. Wireless Insite runs the simulations with given configurations and saves the requested output inside a folder. The script repeats this procedure for a certain number of simulations defined by us. Finally, we use MATLAB to post-process and create the channel frequency responses from the output folders and save them in a database. The roles of components can be summarized as follows:

\begin{figure*}[t]

 \center

  \includegraphics[width=\textwidth]{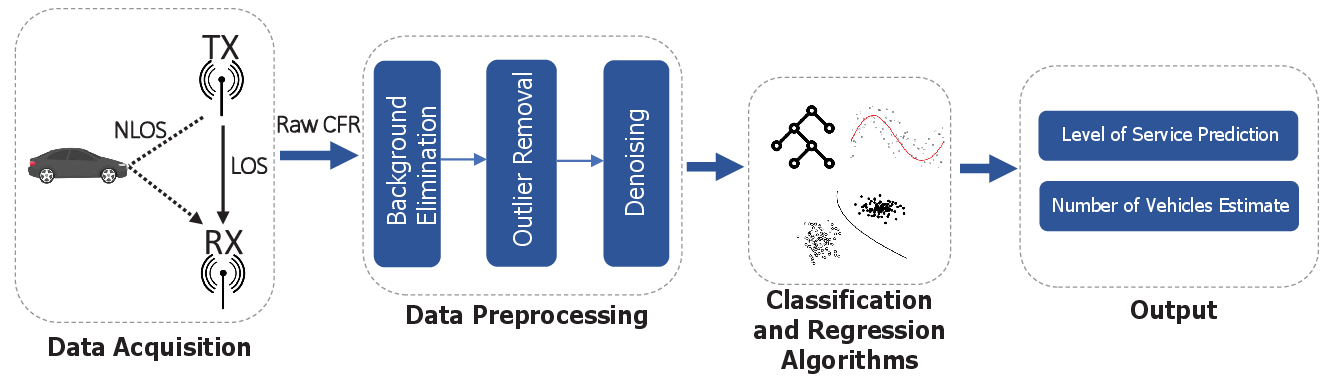}

  \caption{\label{architecture}Architecture overview of the system.}

  \label{AAA}

\end{figure*}

 \begin{itemize}
      \item Traffic Simulator
        \begin{itemize}
            \item Simulating traffic conditions for given mobility parameters.
            \item Providing the positions of the vehicles.
        \end{itemize}
     \item Ray-Tracing Simulator
        \begin{itemize}
            \item Specifying the radio propagation environment with a configuration file.
            \item Simulating the wireless propagation according to the positions of the vehicles.
            \item Saving propagation information for each scenario.
        \end{itemize}
     \item Script
     \begin{itemize}
         \item Running the traffic simulator and obtaining the attitudes of the vehicles.
         \item Modifying the configuration files of the ray-tracing simulator.
          \item Running the ray-tracing simulator via the command line.

     \end{itemize}
     
 \end{itemize}

\subsection{Data Preprocessing}

Within the field of machine learning, data quality is a significant consideration. Data preprocessing is an essential procedure to improve the quality of data and the outcomes of the inference. Fig. \ref{architecture} shows how the data obtained are preprocessed before being fed to machine learning algorithms. In the following sections, we introduce the steps of CFR data preprocessing.

\subsection*{\textbf{Background Elimination}}

CFR does not only embody the reflections from the vehicles but also embodies the reflections from the static environment that should not be learned in machine learning algorithms. The static reflections might cause over-fitting, and degrade the generalization performance of the learning algorithms. Therefore, we need to remove the effect of the static environment from the CFR.

To this end, we estimate a sequence of CFR vectors when there is no vehicle on the road and calculate the average of the CFR vectors $\textbf{H}_b$ as the background CFR. We remove it from the CFR measurements as follows:
 $$\bar{\textbf{H}}_k= \textbf{H}_k - \textbf{H}_b $$
where $\bar{\textbf{H}}_k$ is the CFR of $k^{th}$ frame after the background elimination. This operation ensures that $\bar{\textbf{H}}_k$ preserves only the dynamic reflections shaped by the vehicles.

\subsection*{\textbf{Outlier Removal}}

 Outlier removal is an important step since outliers could affect traffic inference performance. Thus, outliers should be sifted out before further data processing. The purpose of outlier removal is to eliminate and replace outliers with their expected values.

To this end, linear filters are sometimes used for eliminating outliers but it is observed that the linear filters are generally ineffective in this regard and effective outlier removal filters are necessarily nonlinear\cite{outlier_removal}. Further, outlier removal is different from signal filtering. Signal filtering not only removes outliers but also changes the data structure by reducing the data variations. In this regard, outlier removal is more difficult than filtering since it tries to preserve data structure while removing outliers.

We utilize a Hampel filter, obtained by applying the Hampel identifier to a moving data window. In detail, given a sequence $x_1, x_2, x_3, \dots, x_n$ and a sliding window of length $w$, we can define the local median and the standard deviation as follows: 

 \begin{itemize}
     \item $m_i=\text{median} (x_{i-w}, x_{i-w+1}, \dots, x_{i+w-1}, x_{i+w})$
     \item $\sigma_i= \kappa \text{median}(|x_{i-w}-m_i|,\dots,|x_{i+w}-m_i|)$, where $\kappa=\frac{1}{\sqrt{2} \mathit{erf}^{-1}(1/2)}\approx 1.4826$.
\end{itemize}
where $m_i$ and $\sigma_i$ are the local median and the standard deviation. The quantity $\sigma_i / \kappa$ is known as the median absolute deviation (MAD). If a sample $x_i$ is such that

\begin{equation}
    \mid x_i - m_i \mid > n \sigma_i
\end{equation}
for a given threshold $n$, then the Hampel identifier declares $x_i$ an outlier and replaces it with $m_i$.

\begin{figure}[!h]
\includegraphics[height=0.21\textheight]{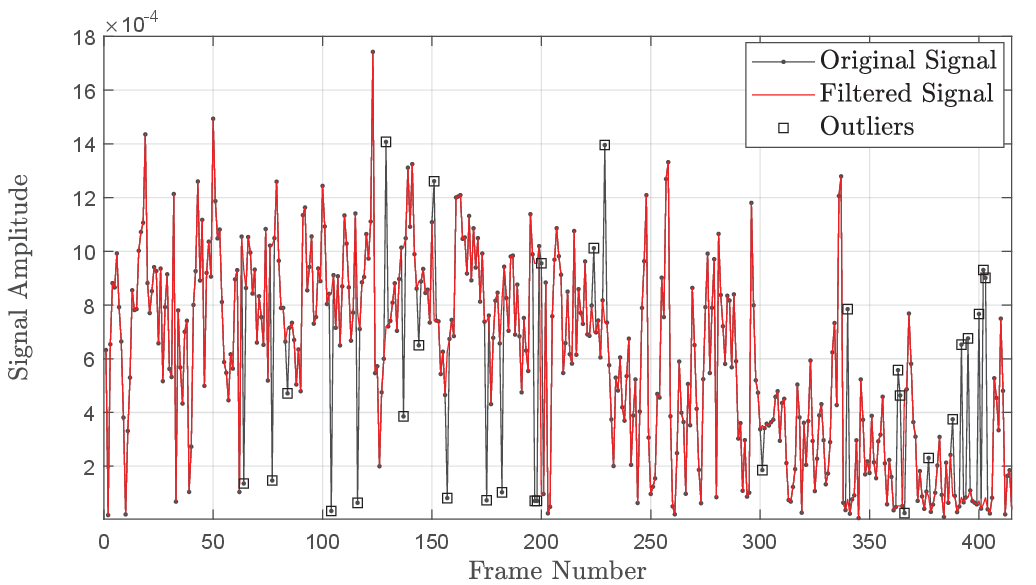}
\centering
\caption{\label{outlier}Original and outlier filtered signals.}
\end{figure}

Fig. \ref{outlier} shows the waveform of one subcarrier and the Hampel filtered version of it. The Hampel identifier adopted here uses the window size of 5 and the threshold $n = 3$. 

\subsection*{\textbf{Denoising}}

Internal state transitions (e.g. transmission strength changes, rate changes) in communication systems, electromagnetic interference, and thermal noise can be listed as the main source of noise in CFR samples. The noises in the raw CFR samples should be wiped out to avoid unnecessary complexity in the learning models and improve the performance of the algorithms. Also, the signal strength variation over distances of the order of the carrier wavelength, due to constructive and destructive interference of multipath components, should be smoothed away to increase the performance of the approach. Since the signal fluctuation created by vehicular activities has low-frequency components, we can adopt a low-pass filter to eliminate the noise and multipath interference in CFR.

The filter to be used should not introduce a large delay to capture the exact time of the events and not distort the signal characteristics. We contend that it is not convenient to utilize traditional filters (e.g., the Butterworth and Chebyshev filters). Specifically, IIR filters present an undesired phase shift (delay) into the filtered signal which varies with the frequency of the signal. This delay can be prevented only if the complete signal is known in advance by using zero-phase filtering techniques which is impossible in real-time measurements. They also soften the rising/falling edges appeared in the signals, which are critical for traffic inference.

\begin{figure}[!h]
\includegraphics[width=0.49\textwidth]{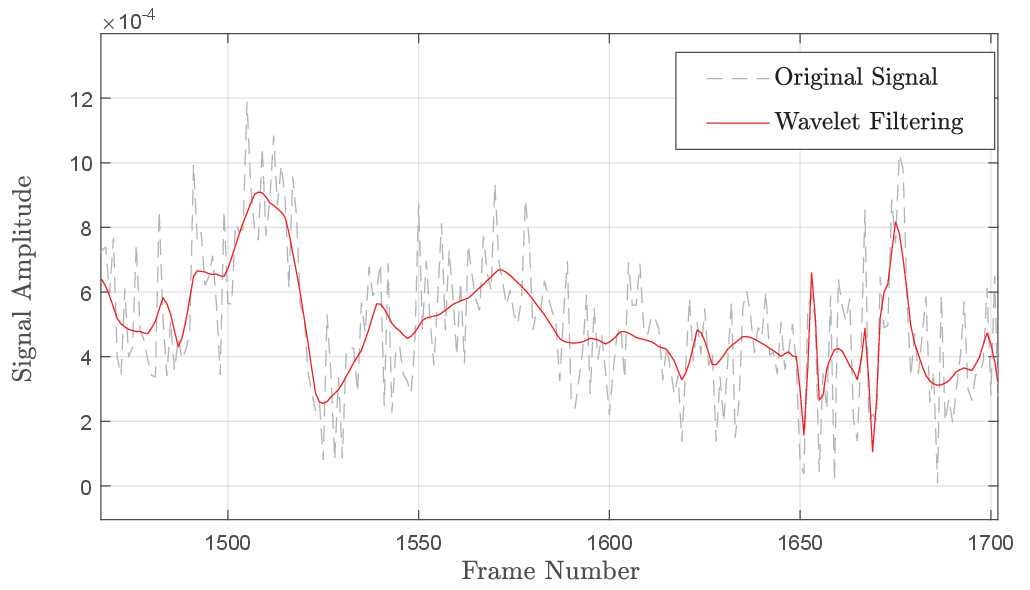}
\centering
\caption{\label{filtering}Performance of the wavelet filter on the signal.}
\end{figure}

We utilize the wavelet filter\cite{wavelet_denoising} since it does not only smooth away the signal but also successfully preserves the sharp transitions.  Fig. \ref{filtering} shows the performance of the wavelet filter on a sub-carrier. The filtering is controlled by the selection of wavelet type and the decomposition level. The higher decomposition level means a lower frequency divider between the signal and noise. Specifically, we employ four levels ‘sym4’ wavelet transform on each sub-carrier signal with the decomposition level of 9. We observe that the wavelet filter captures the abrupt changes well while smoothing away the signal.

\subsection{Machine Learning Algorithms}\label{Features_Section}

In this work, we use SciPy\cite{scipy}, an open-source Python library used for scientific computing, and scikit-learn\cite{scikit-learn} to construct our machine learning framework. Especially, scikit-learn features many machine learning algorithms for classification, regression, and more. In this paper, we use ensemble learning algorithms\cite{sagi2018ensemble} (extremely randomized trees, gradient boosting, random forest), support vector machine (SVM), and k-nearest neighbors (KNN) algorithms to train learning models for classification and regression purposes. The hyperparameter optimization is performed with the grid search method to find the best parameters of the models. The parameters used for each data set can be found at \cite{dataset}.

\subsection*{\textbf{Algorithm Evaluation}}

We use a stratified k-fold cross-validation approach to evaluate the performance of the different algorithms. The approach is similar to k-fold cross-validation. It divides the set of samples into $k$ groups, or \textit{folds}, of approximately equal size and uses $k-1$ folds for the training and the remaining fold for the validation. This procedure is repeated $k$ times and a different group of the samples is used as a validation set each time. In stratified k-fold cross-validation, the folds are formed in a way that each fold contains approximately the same proportion of predictor labels as the original data set to minimize the bias inherited from the random sampling. The stratified k-fold cross-validation estimate of a metric ($CV_{Metric}$) is computed by taking the average of the metric over k folds as:

\begin{equation*}
     \mathit{CV_{Metric}}=\frac{1}{k} \sum_{i=1}^k \text{Metric}_i   
\end{equation*}

%% file: performance.tex
\section{Performance Evaluation}

\subsection{Performance Metrics}

 In this work, the accuracy of the algorithms, the area under the receiver operating characteristic curve (AUC) \cite{bradley1997use}, and the macro-averaged $F_1$ score are employed to evaluate the performance of the level of service prediction problem. The receiver operating characteristic (ROC) curve is plotted with the true positive rate (TPR) against the false positive rate (FPR). Although the receiver operating characteristic (ROC) curve is typically used in the binary classification problem, we adapt it for multi-class classification with the one-vs-all approach and averaging techniques like macro averaging. For each class $i$, the TPR and FPR can be calculated as
 $$\mathit{TPR}_i = \frac{\mathit{TP}_i}{\mathit{TP}_i+\mathit{FN}_i} ,\mathit{FPR}_i = \frac{\mathit{FP}_i}{\mathit{FP}_i+\mathit{TN}_i}  $$
 The macro-averaged TPR and FPR can be calculated as
  $$\mathit{TPR}_{macro} = \frac{\sum _{i=1}^C \mathit{TPR}_i}{k} ,\mathit{FPR}_{macro} = \frac{\sum _{i=1}^C \mathit{FPR}_i}{k}  $$
 where C is the number of classes, which is 6 for the level of service prediction. So, the macro-averaged ROC can be obtained by plotting $\mathit{TPR}_{macro}$ against $\mathit{FPR}_{macro}$ for different threshold values. Here, AUC indicates the performance of the classifier independent of the threshold value and helps us evaluate how well the probabilities from the positive class are separated from the negative class since we have a balanced dataset. Macro-averaged $F_1$ score maintains a balance between the precision and the recall to measure the model's accuracy and it is calculated as the harmonic mean of the macro-averaged recall ($R_{macro}$) and macro-averaged precision ($P_{macro}$) as:

$$F_{1-macro} = \frac{\text{2} P_{macro}R_{macro}}{P_{macro}+R_{macro}}$$
where
  $$P_{macro} = \frac{1}{C}\sum _{i=1}^C \frac{\mathit{TP}_i}{\mathit{TP}_i+ \mathit{FP}_i} ,R_{macro} = \frac{1}{C}\sum _{i=1}^C \frac{\mathit{TP}_i}{\mathit{TP}_i+ \mathit{FN}_i}  $$

We report the mean absolute error (MAE), weighted mean absolute percentage error (WMAPE) as a scale-independent metric, and Pearson correlation coefficient between estimated and actual values as performance metrics for the number of vehicles estimation problem. For $M$ number of samples of X and Y, they are defined as
 $$\mathit{MAE}=\frac{1}{M}\sum^N_{i=1}|x_i - y_i|$$
\begin{equation*}
    \mathit{WMAPE}=\frac{\frac{1}{M}\sum_{i=1}^M|x_i - y_i|}{\frac{1}{M}\sum_{i=1}^M x_i}
\end{equation*}
Pearson correlation coefficient between X and Y ($\rho_{xy}$): $$\rho_{xy} = \frac{\sum_{i=1}^M (x_i-\bar{x})(y_i-\bar{y})}{\sqrt{\sum_{i=1}^M(x_i-\bar{x})^2}\sqrt{\sum_{i=1}^M(y_i-\bar{y})^2}}$$
where $x_i$ and $y_i$ are the actual and estimated value of a sample point, respectively. $\bar{x}$ and $\bar{y}$ are the sample mean of the actual and estimated values. The Pearson correlation coefficient has a value between +1 and -1. The higher correlation between the predicted and actual values implies a better fit of the model to the data.

\subsection{Simulation Setup}

\begin{figure}[b]
\includegraphics[width=0.6\columnwidth]{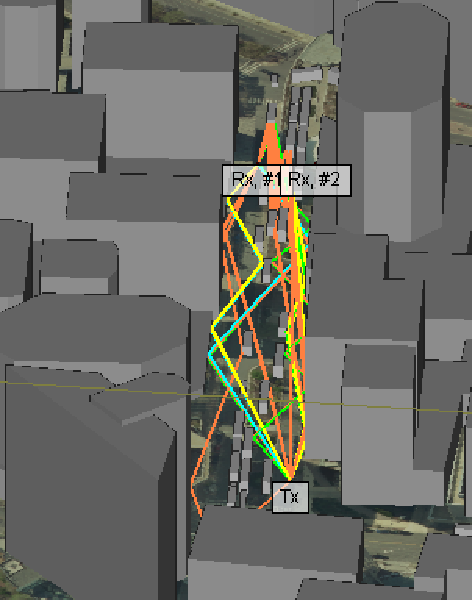}
\centering
\caption{\label{WI} Four-lane road simulation in an urban canyon scenario.}
\end{figure}

Fig. \ref{WI} shows the urban canyon scenario that corresponds to a region in Virginia. The ray-tracing simulator simulates the wireless channel on the four-lane road under different traffic conditions. Two receivers are placed to capture signals transmitted from a transmitter at an intersection that represents a roadside unit. The gray rectangular objects in Fig. \ref{WI} correspond to the vehicles on the road.

\begin{figure}[!b]
\includegraphics[width=0.45\textwidth]{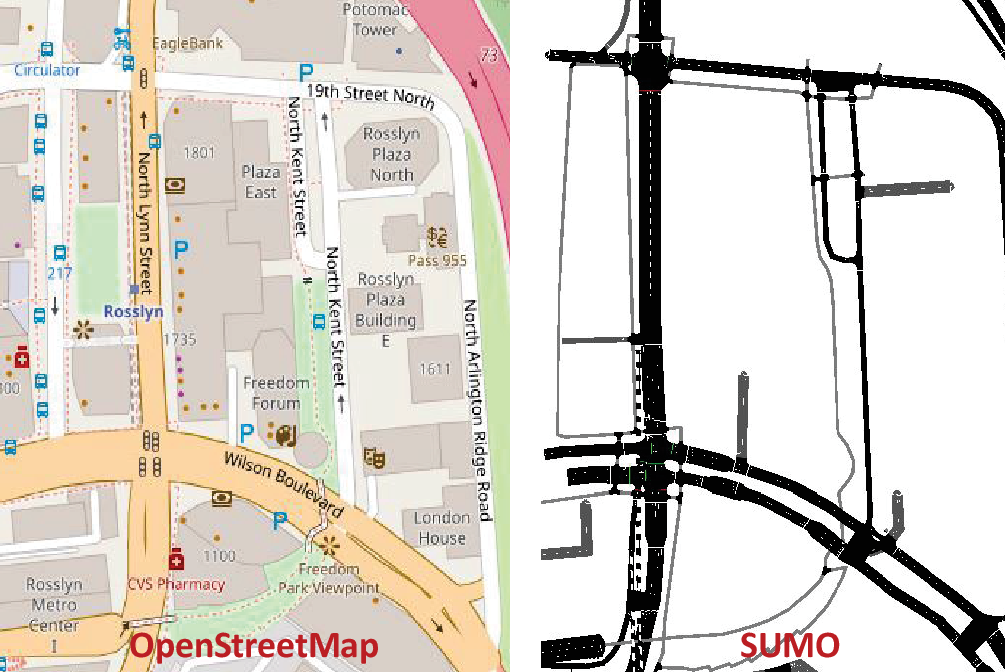}

\caption{\label{Rosslyn} The road simulated in the ray-tracing simulator is converted to the SUMO network from OpenStreetMap (OSM).}
\end{figure}

We import the real-world map of the region from OpenStreetMap (OSM) \cite{OpenStreetMap} and convert OSM files to the SUMO road network format using the network converter tool of SUMO as shown in Fig. \ref{Rosslyn}. Three types of vehicles: car ($1.80$ m$\times 4.60$m$\times 1.60$m), bus ($2.40$m$\times 9.00$m$\times 3.20$m) and trucks ($2.50$m$\times 12.00$m$\times 4.30$m) are simulated in SUMO. We change the probability of injecting a vehicle to the SUMO network to simulate the different traffic conditions on a road segment.

The material decisions of buildings, terrain, and vehicles are vital for simulating realistic simulations. Wireless Insite has a material database that consists of common buildings, terrain, and a few generic material types. We choose the materials according to the ITU (International Telecommunication Union) recommendations and Table \ref{WI_Parameter} shows the details of the ray-tracing parameters.

\begin{table}[h]
\centering
\caption{\label{WI_Parameter} The ray-tracing simulator parameters}
\begin{tabular}{cc}
\cline{1-2} \\[-1em]
\multicolumn{2}{|c|}{{\color[HTML]{000000} \textbf{Wireless Insite parameters}}}   \\ \cline{1-2} \\[-1em]
\multicolumn{1}{|c|}{Propagation Model}                 & \multicolumn{1}{c|}{X3D}                        \\ \cline{1-2} \\[-1em]
\multicolumn{1}{|c|}{Total Number of Rays}                 & \multicolumn{1}{c|}{25}                        \\ \cline{1-2} \\[-1em]
\multicolumn{1}{|c|}{Building Material}                 & \multicolumn{1}{c|}{ITU Layered drywall 5GHz}    \\ \cline{1-2} \\[-1em]
\multicolumn{1}{|c|}{Terrain Material}                  &  \multicolumn{1}{c|}{Asphalt}                     \\ \cline{1-2} \\[-1em]
\multicolumn{1}{|c|}{Vehicle Material}                  & \multicolumn{1}{c|}{Metal}                       \\ \cline{1-2} \\[-1em]
\multicolumn{1}{|c|}{Antenna}                           & \multicolumn{1}{c|}{Half-wave dipole}            \\ \cline{1-2} \\[-1em]
\multicolumn{1}{|c|}{Transmit Power ($P_{\text{Total}}$)}                         & \multicolumn{1}{c|}{30 dBm}                      \\ \cline{1-2} \\[-1em] 
\multicolumn{1}{|c|}{Tx-Rx Antenna Height}                         & \multicolumn{1}{c|}{2 meters}                      \\ \cline{1-2} \\[-1em] 
\multicolumn{1}{|c|}{Carrier Frequency}                 & \multicolumn{1}{c|}{5.9 GHz}                    \\ \cline{1-2} \\[-1em] 
\multicolumn{1}{|c|}{Bandwidth (B)}                         & \multicolumn{1}{c|}{10 MHz}                      \\ \cline{1-2} \\[-1em]

\multicolumn{1}{l}{}                                    & \multicolumn{1}{l}{}                            
\end{tabular}
\centering

\end{table}

\subsection*{\textbf{Simulation Results}}\label{simulation_result_sec}

We have created a data set that includes 12,000 simulations under different traffic conditions using the approach described in Section \ref{approach_section} and performed the stratified 10-fold cross-validation on the data set to obtain the results. For each simulation, we obtain the number of vehicles between the transmitter and receivers from SUMO. Next, we calculate the traffic density in terms of passenger cars per kilometer per lane from the number of vehicles. We observe that the density ranges between 0 and 40 pc/km/lane, and we convert the density values to the corresponding level of service, from A to F, according to Table \ref{los}. We have created a balanced data set that includes 2000 samples from each level of service.

\begin{figure*}[!h]
     \centering
     \begin{subfigure}[b]{\columnwidth}
         \centering
         \includegraphics[width=0.97\columnwidth]{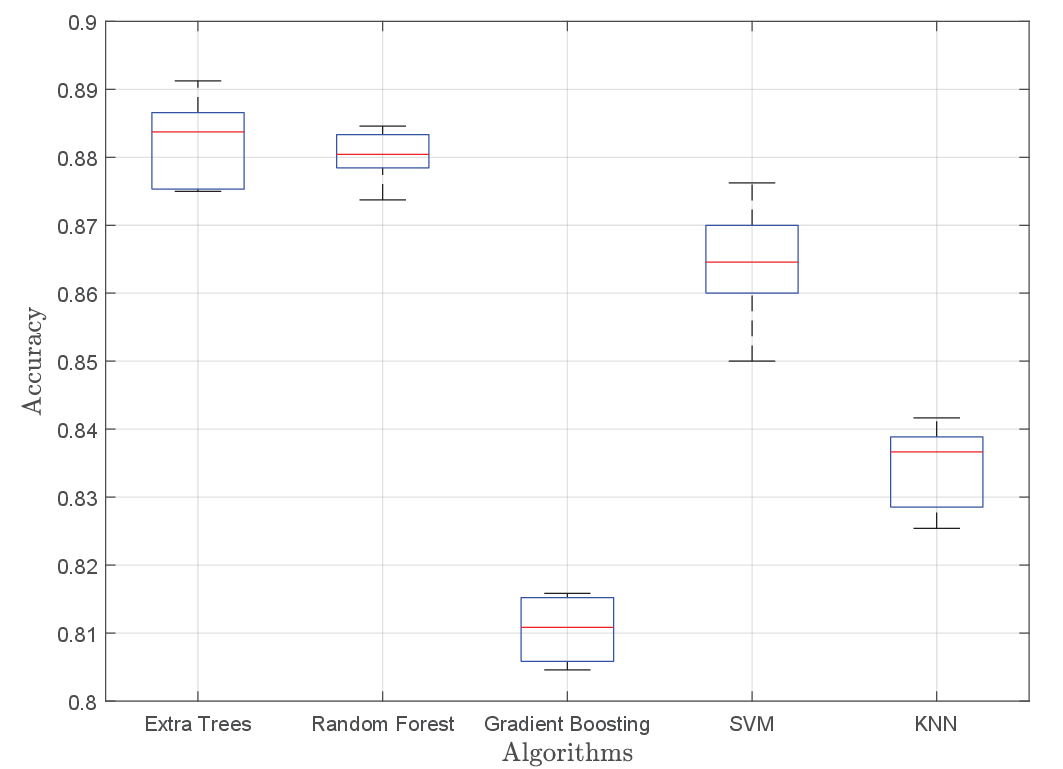}
         \caption{}
     \end{subfigure}
     \hfill
     \begin{subfigure}[b]{\columnwidth}
         \centering
         \includegraphics[width=0.98\columnwidth]{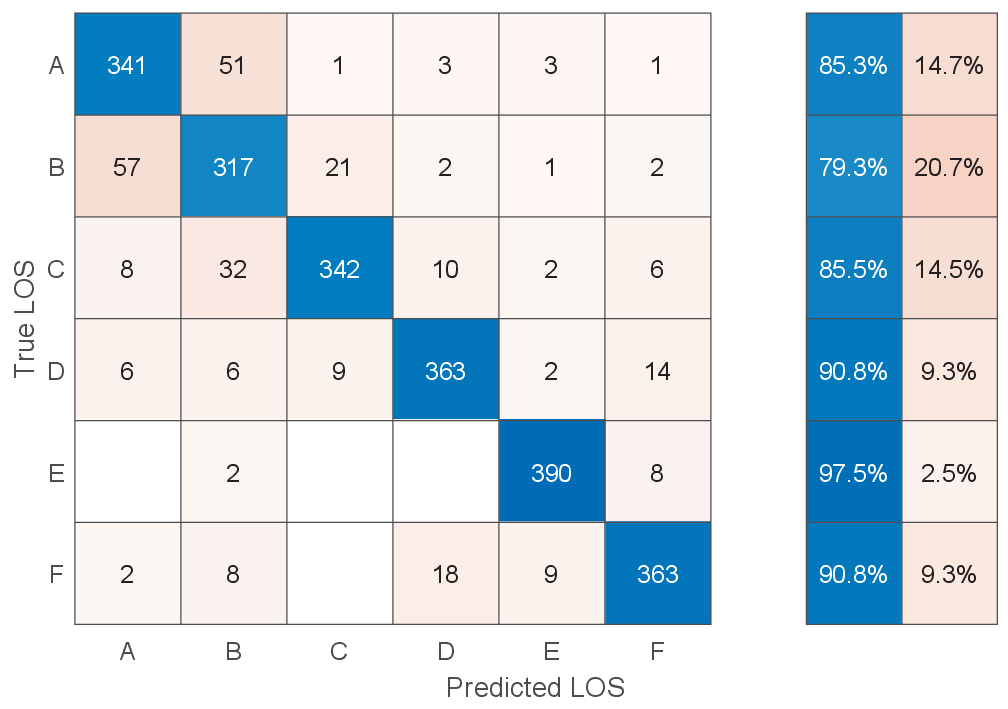}
         \caption{}
     \end{subfigure}
        \caption{a) Box-plot distributions of the classification accuracy resulting from the 10-fold cross-validation of the algorithms. b) Confusion matrix of the extra randomized trees algorithm. }
        \label{box}
\end{figure*}

Table \ref{simulation_classification} shows the accuracy, macro-AUC, and macro-F1 results for the LOS prediction problem. We achieve the best average accuracy of 88.5\% with the extremely randomized trees algorithm. Fig. \ref{box}a shows the distribution of 10-fold cross-validation classification results on a box plot. Fig. \ref{box}b shows the confusion matrix for the extremely randomized trees algorithm. We notice that the mislabeled levels are higher for LOS B-C compared to the other levels. It is an anticipated result considering the narrow boundaries of these levels.

\begin{table}[b]
\caption{\label{simulation_classification} LOS prediction results on the simulation data.}
\begin{tabular}{|c|c|c|c|}
\hline

\textbf{Algorithm} & \textbf{Accuracy} & \multicolumn{1}{l|}{\textbf{Macro-AUC}} & \boldmath{\textbf{Macro-$F_1$}}\\ \hline
Extra Trees        &  $88.5\%$           & 0.98    & 0.88                                                   \\ \hline
Random Forest      & $87.9\%$           & 0.98   & 0.88                                                   \\ \hline
Gradient Boosting  &  $80.7\% $          & 0.95    & 0.80                                                  \\ \hline
SVM                &  $86.7 \%$           & 0.96 & 0.86                                                     \\ \hline
KNN                & $83.5 \%$           & 0.96   & 0.83                                                   \\ \hline

\end{tabular}
\centering

\end{table}

\begin{figure}[!h]
\includegraphics[width=0.46\textwidth]{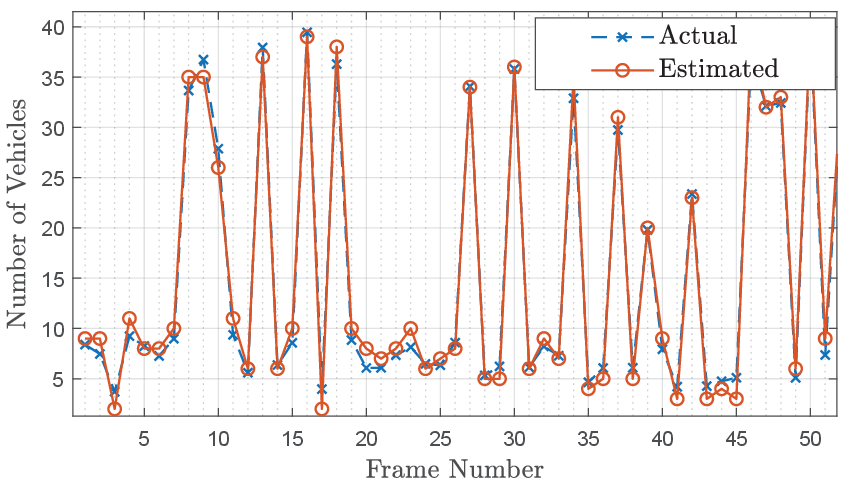}
\centering
\caption{\label{simulation_regression_result} The actual and estimated number of vehicles for 50 frames in the simulation.}
\end{figure}

Afterward, the number of vehicles on four lanes between the transmitter and receivers in Fig. \ref{WI} is estimated. Instead of classification, the algorithms now are used to build a regression model that maps the CFR values observed at two receivers to the number of vehicle estimates. Table \ref{simulation_regression} shows MAE, WMAPE, and the correlation coefficients of the algorithms. We again observe that the extremely randomized trees algorithm outperforms others with a mean absolute error of 2.21 vehicles and a WMAPE of 13.8\%. Fig. \ref{simulation_regression_result} shows the actual and estimated number of vehicles for 50 frames obtained using the extremely randomized trees algorithm.

\begin{table}[t]
\caption{\label{simulation_regression}Number of vehicles estimation results on the simulation data.}
\begin{tabular}{|c|c|c|c|}
\hline

{ \textbf{Algorithm}} & { \textbf{MAE}} & { \textbf{WMAPE}} & { \textbf{Correlation Coef.}} \\ \hline
Extra Trees                               & 2.21                                & 13.8\%                                 & 0.96                                              \\ \hline
Random Forest                             & 2.31                                & 14.4\%                                & 0.96                                              \\ \hline
Gradient Boosting                         & 2.60                               & 15.9\%                                 & 0.95                                             \\ \hline
SVR                                       & 2.70                                & 16.3\%                                  & 0.94                                              \\ \hline
KNN                                       & 2.48                                & 15.4\%                                 & 0.94                                               \\ \hline

\end{tabular}
\centering

\end{table}

\subsubsection*{\textbf{Effect of Signal-to-Noise Ratio (SNR)}} The signal quality is an important consideration for machine learning algorithms. Especially, the noise in data can prevent knowledge extraction from the data. The noise can impair the models trained with such data since the algorithm can interpret the data noise as a pattern and try to generalize from it \cite{garcia2015dealing}. We evaluate how the signal quality (i.e., signal-to-noise ratio) affects our results by varying the transmitter power level. The SNR of $k^{th}$ frame at a receiver can be defined as:
$$\text{SNR}_k = \frac{P_R}{\sigma_W^2} = \frac{P_{T}\|\textbf{H}_k\|^2}{k_BTN_FB}$$
where $P_R$ is the total received power, and $P_T$ is the transmitter power per subcarrier which is equal to $P_{\text{Total}}/N$ with uniform power allocation across the subcarriers. $k_B=1.38 \times 10^{-23} J/K$ is the Boltzmann's constant, $T$ is the temperature in Kelvin (K), $N_F$ is the receiver noise figure, and $B$ is the total bandwidth. 

To evaluate the performance of the algorithms with different SNR levels, we decrease the total transmit power to $20$ dBm from $30$ dBm which results in a $10$ dB decrease in SNR. The instantaneous SNR ranges from $-35.3$ dB to $50.6$ dB at the first receive antenna with a mean value of 19.7 dB, by using $P_{\text{Total}}=20$ dBm, $B=10$ MHz, and $T=300$ K, $N_F=2$. Table \ref{simulation_classification_20dbm} shows the accuracy, macro-AUC, and macro-F1 results for the LOS prediction problem with 20 dBm transmitter power. We observe that the classification accuracies decrease by 1-5.2\% compared to Table \ref{simulation_classification}. Since SVM is sensitive to noisy data \cite{atla2011sensitivity} \cite{sabzekar2021robust}, especially with a low-bias kernel, the performance of SVM weakens significantly among the other algorithms. 

\begin{figure}[!t]
\includegraphics[width=0.5\textwidth]{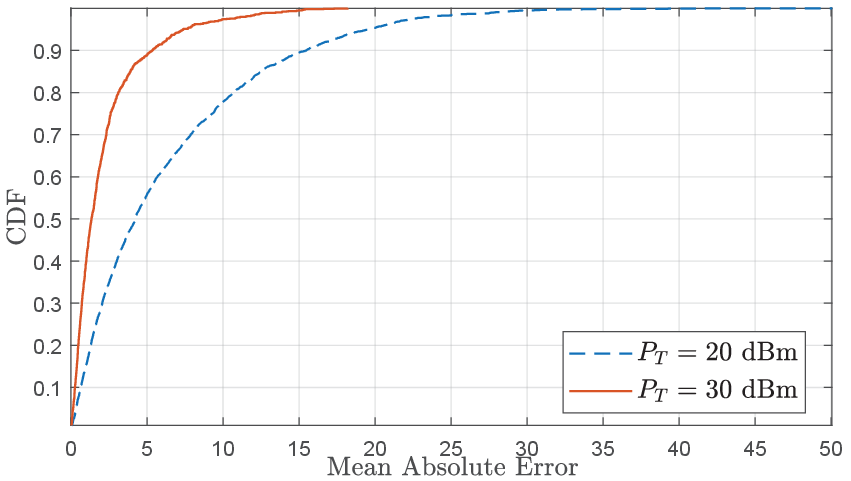}
\centering
\caption{\label{SNR_MAE} The CDF of the mean absolute error under different transmit power levels.}
\end{figure}

Table \ref{simulation_regression_20dbm} shows the number of vehicles estimation results with 20 dBm transmit power. Again, we observe a performance degradation from the results in Table \ref{simulation_regression}. Fig. \ref{SNR_MAE} shows the cumulative distribution function (CDF) of MAE with the extremely randomized trees algorithm. We note the right shift of the curve that indicates the increasing error when we decrease the transmit power.

\begin{table}[!b]
\caption{\label{simulation_classification_20dbm} LOS prediction results with 20 dBm transmit power.}
\begin{tabular}{|c|c|c|c|}
\hline

\textbf{Algorithm} & \textbf{Accuracy} & \multicolumn{1}{l|}{\textbf{Macro-AUC}} & \boldmath{\textbf{Macro-$F_1$}}\\ \hline
Extra Trees        &  $87.1\%$           & 0.98    & 0.87                                                   \\ \hline
Random Forest      & $86.9\%$           & 0.98   & 0.87                                                   \\ \hline
Gradient Boosting  &  $79.7\% $          & 0.95    & 0.79                                                  \\ \hline
SVM                &  $81.5 \%$           & 0.87 & 0.81                                                     \\ \hline
KNN                & $81.4 \%$           & 0.94   & 0.81                                                   \\ \hline

\end{tabular}
\centering

\end{table}

\begin{table}[!t]
\caption{\label{simulation_regression_20dbm} Number of vehicles estimation results with 20 dBm transmit power.}
\begin{tabular}{|c|c|c|c|}
\hline

{ \textbf{Algorithm}} & { \textbf{MAE}} & { \textbf{WMAPE}} & { \textbf{Correlation Coef.}} \\ \hline
Extra Trees                               & 2.31                                & 14.53\%                                 & 0.96                                              \\ \hline
Random Forest                             & 2.42                                & 15.3\%                                & 0.95                                              \\ \hline
Gradient Boosting                         & 2.69                               & 16.9\%                                 & 0.95                                             \\ \hline
SVR                                       & 4.92                                & 31.1\%                                  & 0.75                                              \\ \hline
KNN                                       & 2.84                                & 17.9\%                                 & 0.92                                               \\ \hline

\end{tabular}
\centering

\end{table}

\subsubsection*{\textbf{Performance with a single antenna}}
We also evaluate the single antenna scenario in which it is not possible to obtain signals from multiple receivers. To this end, the single antenna scenario where only Rx\#1 in Fig. \ref{WI} is used for data creation by disabling Rx\#2 in the simulations. Table \ref{classification_single} and Table \ref{regression_single} show the performance of the algorithms with the single antenna for the level of service prediction and the number of vehicles estimation problems, respectively. When we compare the results with Table \ref{simulation_classification} and \ref{simulation_regression}, we observe that the results worsen when we use a single antenna. Specifically, the classification accuracies drop by 2.6-3.4\%,  and WMAPEs of the algorithms increase by 2.9-5\%  with the single antenna. Hence, we observe a significant improvement in the performance by utilizing a second antenna. With this motivation, we evaluate the performance with a third and a fourth antenna in the simulations. However, the performance improvement is not significant and we observe a diminishing return while increasing the number of antennas.

\begin{table}[!h]
\caption{\label{classification_single} LOS prediction results with the single antenna.}
\begin{tabular}{|c|c|c|c|}
\hline

\textbf{Algorithm} & \textbf{Accuracy} & \multicolumn{1}{l|}{\textbf{Macro-AUC}} & \boldmath{\textbf{Macro-$F_1$}}\\ \hline
Extra Trees        &  $86.6\%$           & 0.98    & 0.86                                                   \\ \hline
Random Forest      & $86.1\%$           & 0.98   & 0.86                                                    \\ \hline
Gradient Boosting  &  $74.3\% $          & 0.93    & 0.74                                                  \\ \hline
SVM                &  $82.9 \%$           & 0.95 & 0.82                                                     \\ \hline
KNN                & $81.7 \%$           & 0.94   & 0.81                                                   \\ \hline

\end{tabular}
\centering

\end{table}

\begin{table}[!h]
\caption{\label{regression_single} Number of vehicles estimation results with the single antenna.}
\begin{tabular}{|c|c|c|c|}
\hline

{ \textbf{Algorithm}} & { \textbf{MAE}} & { \textbf{WMAPE}} & { \textbf{Correlation Coef.}} \\ \hline
Extra Trees                               & 2.74                                & 17.1\%                                 & 0.93                                              \\ \hline
Random Forest                             & 2.86                                & 17.9\%                                & 0.93                                              \\ \hline
Gradient Boosting                         & 3.16                               & 20.1\%                                 & 0.92                                             \\ \hline
SVR                                       & 3.30                                & 20.5\%                                  & 0.90                                              \\ \hline
KNN                                       & 3.01                                & 18.5\%                                 & 0.92                                               \\ \hline

\end{tabular}
\centering

\end{table}

\subsection{Real-world Experiments}

The feasibility of the proposed approach is tested under a real-world DSRC communication scenario by collecting signal phase and timing (SPaT) messages broadcasted from a roadside unit. In the experiments, a  software-defined radio is used for data acquisition and an action camera recorded the road to obtain the ground truth of the number of vehicles. Our experiments include two types of scenarios: 

\begin{center}
  \begin{enumerate}
  \item Static transmitter-static receiver experiments.
  \item Static transmitter-moving receiver experiments.
\end{enumerate}  
\end{center}

The details of the scenarios and the experimental setup are summarized in the following sections. 

\subsection*{\textbf{Hardware and Software Setup}}

Even though there are commercial IEEE 802.11p modems, they only provide minimal access to the physical layer. For this reason, we prefer working with an open-source software tool, GNU Radio \cite{url:gnu-radio}, and building our receiver prototype using a software-defined radio. X300 of Ettus Research with a TwinRX daughterboard is used for the experiments. We choose the TwinRX daughterboard since it is sufficient for the DSRC spectrum that is allocated from 5.850 to 5.925 GHz with 10 MHz subchannels \cite{DSRC_Stand}. Table \ref{tab:hardware} shows the hardware used in the experiment.

GNURadio is a framework that contains signal processing blocks for software-defined radios. The authors in \cite{bloessl2013ieee} presented the first OFDM receiver for the GNU Radio which supports IEEE 802.11a/g/p and channel bandwidth up to 20MHz. We modified this implementation to enable us to obtain the CFR values of subcarriers and the correct time of each frame reception. The time information is subsequently used to synchronize the frame reception time with the video records. While the receiver provides different algorithms for channel estimation, we use the least-squares channel estimation algorithm to estimate the channel frequency response vector using the long training sequence of the received frames.

\begin{table}[!h]

\centering
\caption{\label{tab:hardware} Hardware used for the experiments.}
\begin{tabular}{|c|c|}

\cline{1-2} \\[-1em]
{ \textbf{Component}} & {\color[HTML]{000000} \textbf{Type}}  \\ \cline{1-2} \\[-1em]
CPU                                                               & Intel Core i7-4720 HQ CPU 2.6GHz                              \\ \cline{1-2} \\[-1em]

USRP                                                              & Ettus X300                                                     \\ \cline{1-2} \\[-1em]
RF Daughter Board                                                 & Ettus TwinRx                                                  \\ \cline{1-2} \\[-1em]
RF Antenna                                                        & VERT 2450                                                      \\ \cline{1-2} \\[-1em]
Camera                                                        & Akaso 12 MP Action Camera  1080p                                                    \\ \cline{1-2} 
\end{tabular}


\end{table}
\subsection*{\textbf{Static Transmitter-Receiver Setup}}

\begin{figure}[!h]
\includegraphics[width=0.42\textwidth]{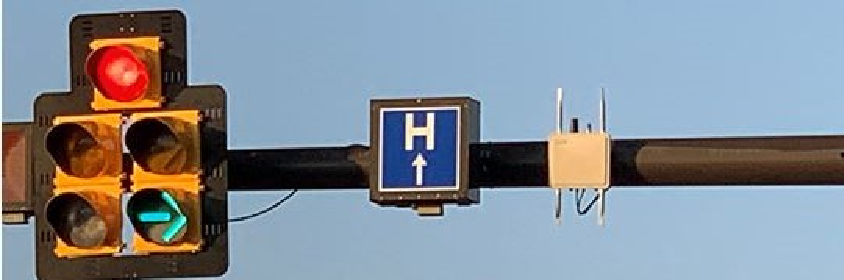}
\centering
\caption{\label{tab:RSU} The roadside unit at the intersection. Photographed by City of Dublin \cite{rsu_pic}.}
\end{figure}

\begin{figure}[!t]
\includegraphics[width=0.45\textwidth]{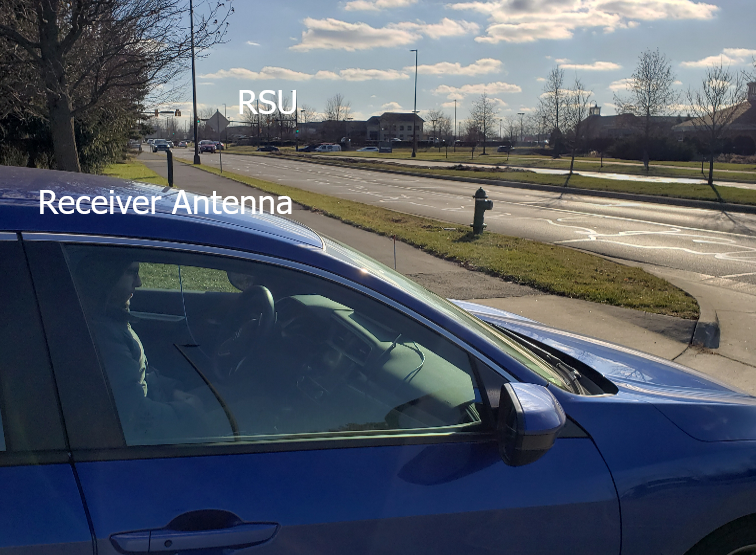}
\centering
\caption{\label{exp_setup}The vehicle collected SPaT messages 120 meters away from the roadside unit.}
\end{figure}

This experiment aims to imitate the simulation setup for a single receiver-transmitter pair and evaluate the performance of our approach in the real world. For this experiment, we have equipped a vehicle with our software-defined radio and experimented on the 33 Smart Mobility Corridor. The corridor is a 35-mile highway test corridor that aims to test real-world autonomous and connected vehicle technologies. To this end, roadside units are placed at intersections to broadcast SPaT messages and other safety messages. Fig. \ref{tab:RSU} shows the roadside unit that is utilized during our experiments. Our experiment setup is shown in Fig. \ref{exp_setup}. 
We have collected SPaT messages from the RSU while our vehicle is parked 120 meters away from the roadside unit. 5700 SPaT messages are collected during the experiment and the vehicles on two lanes close to the parking location are counted from the video record to obtain the actual number of vehicles. The number of vehicles on the road ranges between 0 and 11 which results in a maximum of 46 pc/km/lane on two lanes along the experiment.

\subsection*{\textbf{Static Transmitter-Receiver Results}}

\begin{figure}[!h]
\includegraphics[width=0.50\textwidth,height=0.23\textheight]{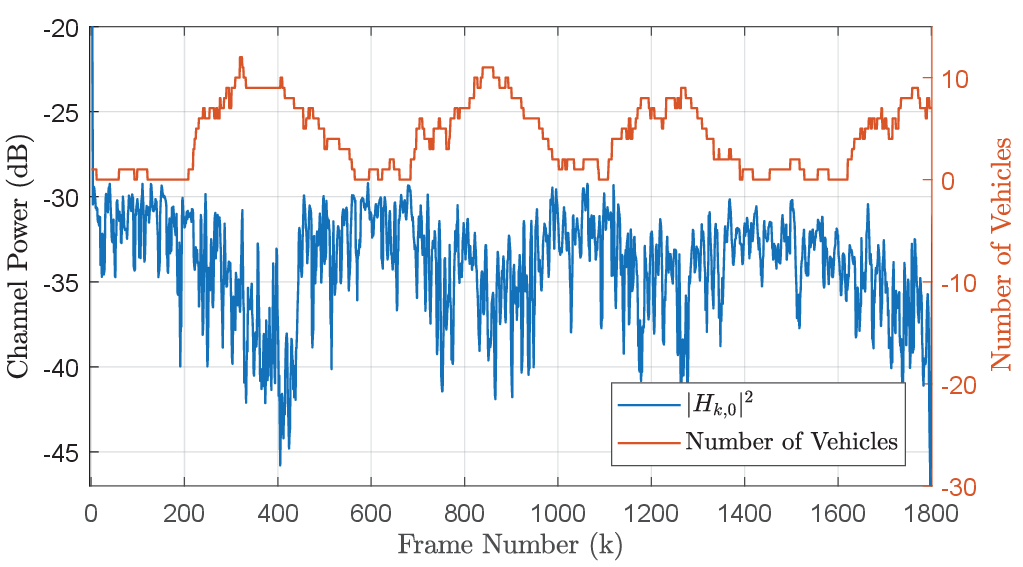}
\centering
\caption{\label{Freq_Response} The power of the first pilot subcarrier and the number of vehicles on the road.}
\end{figure}

We time-stamp each received frame from the RSU to log the time of frame reception. With the help of the timestamps, we determine the number of vehicles at the frame reception time from the video records. Fig. \ref{Freq_Response} shows the power of the first pilot sub-carrier and the number of vehicles. It is seen that there is a correlation between the power level and the number of vehicles on the road.

First, we predict the level of service on the road between the receive antenna and the roadside unit. The stratified 10-fold cross-validation results of the accuracy, macro-AUC, and macro-F1 are shown in Table \ref{experiment_classification}. We obtain accuracy results above 81.7\% and the extremely randomized trees algorithm achieves the best accuracy with 91.8\%.

\begin{figure}[!b]
\includegraphics[width=0.47\textwidth]{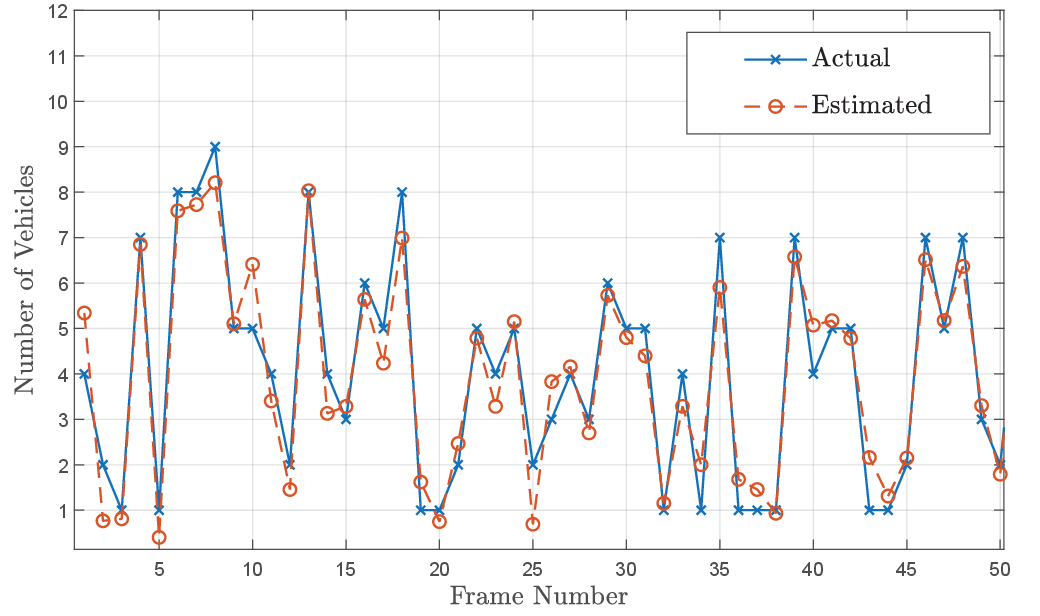}
\centering
\caption{\label{exp_figure}The actual and estimated number of vehicles for 50 frames.}
\end{figure}

Next, we estimate the number of vehicles on two lanes closest to the receive antenna while the number of vehicles on the road ranges between 0 and 11. The results are shown in Table \ref{experiment_regression}. The extremely randomized trees algorithm reaches a mean absolute error of 0.93 vehicles and a WMAPE of 23.4\%. Fig. \ref{exp_figure} shows the estimated and actual values for 50 received frames obtained using the extremely randomized trees algorithm.  
\begin{table}[!t]
\caption{\label{experiment_classification} LOS prediction results on the experimental data.}
\begin{tabular}{|c|c|c|c|}
\hline

\textbf{Algorithm} & \textbf{Accuracy} & \multicolumn{1}{l|}{\textbf{Macro-AUC}} & \boldmath{\textbf{Macro-$F_1$}}\\ \hline
Extra Trees        &  $91.8\%$           & 0.99    & 0.91                                                   \\ \hline
Random Forest      & $91.4\%$           & 0.99   & 0.90                                                    \\ \hline
Gradient Boosting  &  $81.7\% $          & 0.94    & 0.80                                                   \\ \hline
SVM                &  $82.9 \%$           & 0.95 & 0.85                                                     \\ \hline
KNN                & $87.3 \%$           & 0.96   & 0.86                                                   \\ \hline

\end{tabular}
\centering

\end{table}

\begin{table}[!t]
\caption{\label{experiment_regression} Number of vehicles estimation results on the experimental data.}
\begin{tabular}{|c|c|c|c|}
\hline

{ \textbf{Algorithm}} & { \textbf{MAE}} & { \textbf{WMAPE}} & { \textbf{Correlation Coef.}} \\ \hline
Extra Trees                               & 0.93                                & 23.4\%                                 & 0.85                                              \\ \hline
Random Forest                             & 0.98                                & 24.5\%                                & 0.85                                              \\ \hline
Gradient Boosting                         & 1.07                               & 26.8\%                                 & 0.82                                             \\ \hline
SVR                                       & 1.22                               & 32.2\%                                  & 0.80                                              \\ \hline
KNN                                       & 0.97                                & 26.7\%                                 & 0.83                                               \\ \hline

\end{tabular}
\centering

\end{table}

When we compare the simulation and the static transmitter-receiver results on the estimation of the number of vehicles, we observe that the simulation results are better than the experimental results in terms of WMAPE and the correlation coefficient. However, we note that a small error in the experiment results translates into a large percentage error since the actual number of vehicles on the road is mostly close to 0 and WMAPE is sensitive to such numbers. This is one of the reasons we have a larger percentage error as compared to the simulation results.

\subsection*{\textbf{Static Transmitter - Moving Receiver Setup}}

\begin{figure}[b]
\includegraphics[width=0.48\textwidth]{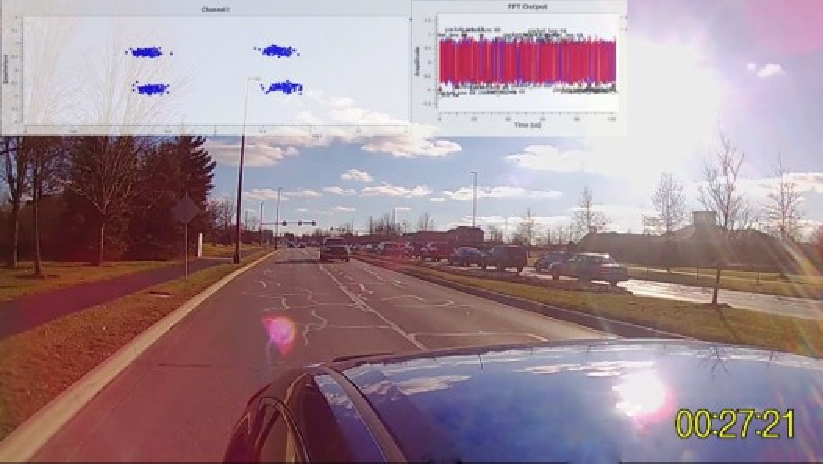}
\centering
\caption{\label{moving}SPaT messages collected from a moving vehicle. The received signal and the constellation diagram are seen at the upper left.}
\end{figure}

After static transmitter-receiver experiments, we have extended our experiments with a more challenging scenario, a moving receiver scenario. With the help of this experiment, we aim to evaluate the performance of our approach when the signals are received from a moving vehicle at a known location. During the experiments, we drive our vehicle in the right lane toward the location where we collected the static data and record the road using the camera on the vehicle. Fig. \ref{moving} shows a caption from the video recorded while we are collecting data along the road. The video can also be found at \cite{dataset}.

\begin{figure*}[t]
     \centering
     \begin{subfigure}[b]{\textwidth}
         \centering
         \includegraphics[width=\textwidth]{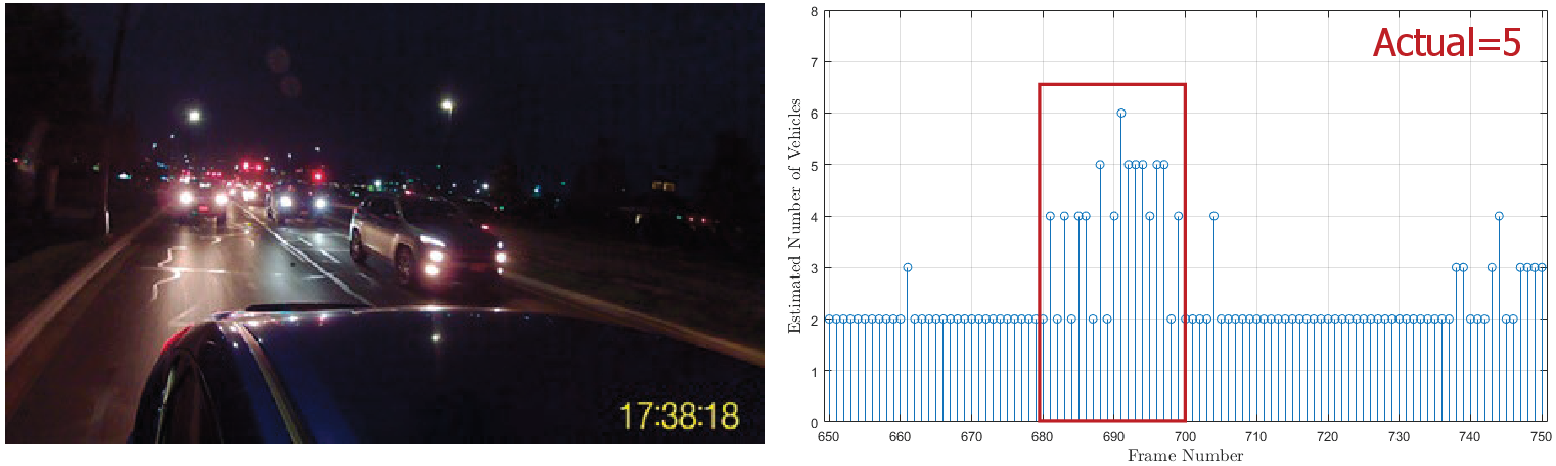}
     \end{subfigure}
     \hfill
     \begin{subfigure}[b]{\textwidth}
         \centering
         \includegraphics[width=\textwidth]{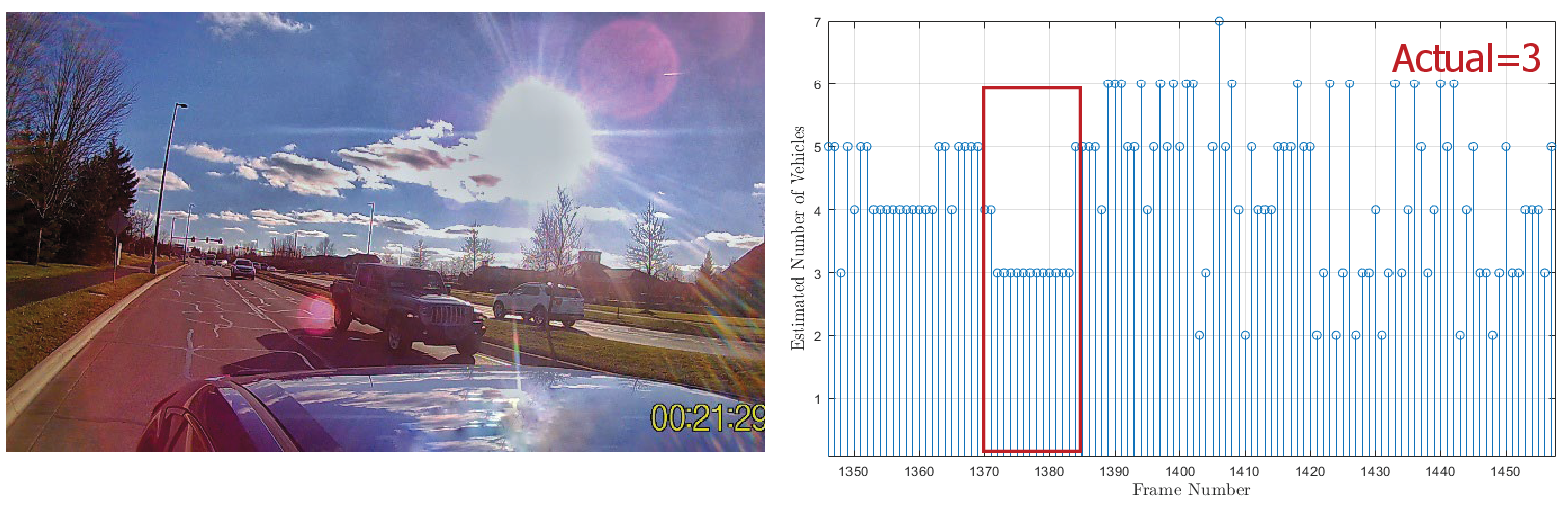}
     \end{subfigure}
        \caption{The results of the two passes using the frames received from the vehicle. }
        \label{moving_test}
\end{figure*}

\subsection*{\textbf{Static Transmitter- Moving Receiver Results}}

Since collecting enough data from a moving vehicle to train a machine learning model may not be feasible for some roads, we propose using models trained with the data collected from a static receiver or simulation data. In this experiment, we use our best model trained with the data obtained from the static receiver scenario to directly estimate the number of vehicles on the road. So, the data obtained from the moving vehicle is merely used for test purposes with the best model.

We drive in the right lane and collect data while driving 5 meters away from the location where we collected the static data. The frames received from the moving vehicle are later used for testing. We have passed on the road 8 times and collected 140 frames in total. In these experiments, we achieve to estimate the number of vehicles with a mean absolute error of 0.95 and a WMAPE of 24\%. Fig. \ref{moving_test} shows the estimated number of vehicles of two passes alongside the video records. We round the estimates to the nearest integer to ease the comparison with the actual values. The frames received in the proximity (under 5 meters away from the static receiver location) are indicated in rectangular frames on the plots. We observe that the performance of the trained model is affected by the location mismatch. Specifically, the estimation error increases with the location mismatch. When we approach the static receiver location, the outputs of the algorithm get closer to the actual number of vehicles as expected.

%% file: conclusion.tex
\section{CONCLUSION}
In this work, we introduce a novel traffic monitoring approach that is based on signal level measurements at a receiver and machine learning techniques. It is a cost-effective and non-intrusive approach. We infer the level of service and the number of vehicles on a roadway using the channel frequency response estimated at a receiver. Since there is no available data, including the traffic conditions and the channel frequency response values, we first create a dataset using a ray-tracing simulator and a traffic simulator. Next, we conduct real-world experiments by collecting DSRC messages broadcast from a roadside unit. Both simulation and experimental results show that the proposed approach is capable of estimating the number of vehicles and predicting the level of service. Our system aims to exploit the infrastructures of vehicular networks, and it doesn't need the deployment of a dedicated device. It can enhance the performance of a current traffic monitoring system when used alongside the system.

\section*{Acknowledgement}

We thank Frank Barickman (NHTSA), John Martin (NHTSA), Sughosh Rao (NHTSA), Gavin Howe (NHTSA), and Keith A. Redmill (OSU) for useful discussions in our meetings.